\documentclass[conference]{IEEEtran}
\usepackage{cite}

\ifCLASSINFOpdf
\else
  \usepackage[dvips]{graphicx}
  \graphicspath{{Figures/}}
\fi
\usepackage{url}

\begin{document}
%
\title{Online classification for time-domain astronomy}

\author{\IEEEauthorblockN{Kitty K. Lo and Tara Murphy}
\IEEEauthorblockA{ CAASTRO and \\
Sydney Institute for Astronomy\\
School of Physics, The University of Sydney\\
NSW, Australia\\
Email: kitty@physics.usyd.edu.au}
\and
\IEEEauthorblockN{Umaa Rebbapragada and Kiri Wagstaff}
\IEEEauthorblockA{Jet Propulsion Laboratory\\
California Institute of Technology\\
Pasadena, California, USA}
}


%

\maketitle

\begin{abstract}
The advent of synoptic sky surveys has spurred the development of techniques for real-time classification of astronomical sources in order to ensure timely follow-up with appropriate instruments.  Previous work has focused on algorithm selection or improved light curve representations, and naively convert light curves into structured feature sets without regard for the time span or phase of the light curves.  In this paper, we highlight the violation of a fundamental machine learning assumption that occurs when archival light curves with long observational time spans are used to train classifiers that are applied to light curves with fewer observations. We propose two solutions to deal with the mismatch in the time spans of training and test light curves.  The first is the use of classifier committees where each classifier is trained on light curves of different observational time spans.  Only the committee member whose training set matches the test light curve time span is invoked for classification.  The second solution uses hierarchical classifiers that are able to predict source types both individually and by sub-group, so that the user can trade-off an earlier, more robust classification with classification granularity.   We test both methods using light curves from the MACHO survey, and demonstrate their usefulness in improving performance over similar methods that naively train on all available archival data.
\end{abstract}



%
\IEEEpeerreviewmaketitle

\section{Introduction}
The advent of next generation optical and radio telescopes such as the Large Synoptic Survey Telescope (LSST) \cite{tyson_large_2003} and the Square Kilometre Array (SKA) \cite{taylor2013} will enable massive wide-field surveys for highly variable or transient astronomical sources. Prompt multi-wavelength follow-up of candidate sources will be critical in achieving the scientific goals of these surveys. A global network of potential follow-up instruments is available, but due to limited resources, only high quality candidates will be selected for follow-up. In the regime of big data surveys, manual inspection of data is no longer possible, therefore automated classification methods must form part of the data pipelines to determine whether a detected candidate is an object of interest that requires follow-up.

The use of machine learning for source type classification of archival light curve data is well-established \cite{eyer_automated_2004,debosscher_automated_2007,sarro_automated_2008-1,kim2011,richards2011,blomme_automated_2011}. Some methods focus on specific events or source types to classify \cite{kim2011}, while others focus on periodic variables \cite{blomme_automated_2011} or a range of source types \cite{debosscher_automated_2007,richards2011}.  All must contend with the challenges inherent in classifying light curve data including the generation of robust training data \cite{long_optimizing_2012}, the search for discriminating features \cite{debosscher_automated_2007,richards2011} and methods that are robust to unevenly sampled data, as well as missing, spurious and sparse observations. Various classification algorithms, such as neural networks, support vector machines and random forests, have been explored. However, algorithmic selection is not the most significant factor in achieving high performance. 

Current surveys that use machine learned classification in their data processing pipelines include the Palomar Transient Factory (PTF) \cite{law_palomar_2009} and the Catalina Real-time Survey (CRTS) \cite{djorgovski_catalina_2011}. PTF employs an automated ``real-bogus'' classification system that identifies true astronomical transient candidates using features extracted from candidates in subtracted images \cite{bloom_towards_2008, brink_using_2012}. To date, the PTF system performs binary classification (real or not), and primarily uses image features rather than light curve information in its decision-making. CRTS has developed a real-time transient detection pipeline that contends with sparse and heterogeneous data sources \cite{djorgovski_towards_2011}, self-updates as observations are received, and makes robust decisions on known classes while potentially discovering unknown sources.

Whether performing real-time or archival classification, the use of machine learning algorithms in the methods cited above necessitates the creation of a structured data set of features from light curves that are arbitrary in time span and out of phase with each other.  After training a classifier on this set, a partially-observed light curve is converted into the same feature characterization and classified.  Theoretically the test examples belong to the same population as the training examples, however the fact that the test light curves are only partially-observed implies that the extracted features may effectively belong to a different statistical distribution.  This violates a basic premise of all supervised learning algorithms that both training and test examples must belong to the same distribution.  A large enough violation of this premise will degrade the performance of the classifier, as it cannot generalize to new test data. To our knowledge, this issue has not  been addressed in previous work. 


%
This paper discusses how to do classification robustly with light curve data arriving in a stream. This type of algorithm is known as an {\it online learning algorithm} and it incorporates new data in the classification model as they become available. There are two novel aspects to our work.  First, in order to handle the mismatch between the light curve time spans in training and test sets, we developed a method that builds a committee of classifiers each trained with light curves of a set time span.  The appropriate committee member that matches the test light in time span is invoked for classification.  We show that this method that can outperform a naive method that trains a single classifier using all available archival data.  Second, we implemented hierarchical classifiers. A non-hierarchical classifier assigns a single label from a set of discrete labels to a test example, while a hierarchical classifier assigns either a single label or a sub-grouping of discrete labels, where sub-groups are pre-defined by the end user in a tree-like structure.  The advantage of a hierarchical classifier is its ability to trade classification granularity against earlier predictions. Given a user-specified desired confidence level, the classifier may not be able to predict down to a single discrete label (e.g., RR Lyrae) but able to predict an informative sub-group of the classes (e.g., variable stars).  This enables the end user of the system to receive some information earlier rather than a more confident prediction much later.  

\section{Data}\label{s_data}

\begin{figure*} 
\begin{center} 
\begin{tabular}{cccc}
    \includegraphics[width=0.4\textwidth]{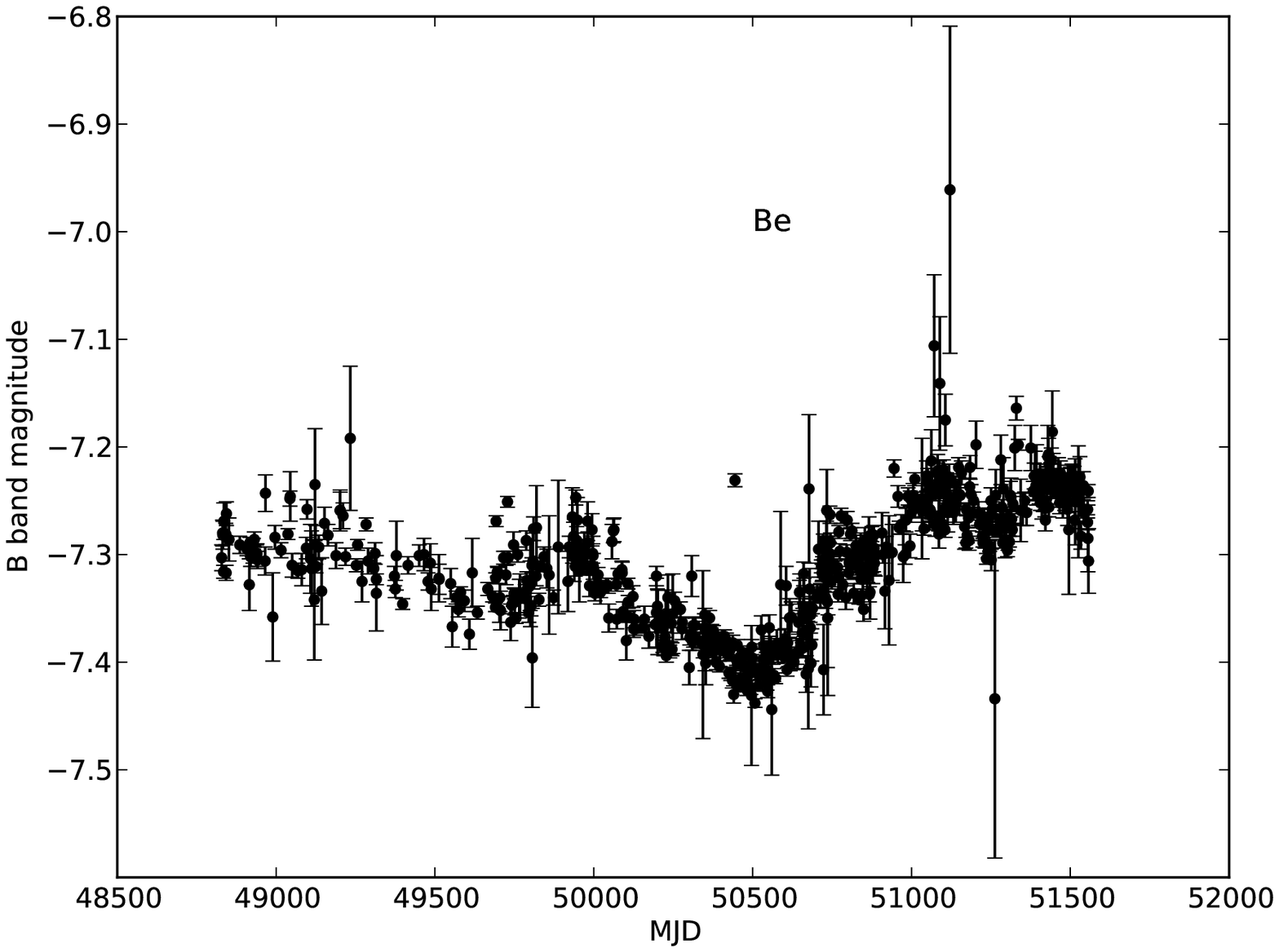}&

    \includegraphics[width=0.4\textwidth]{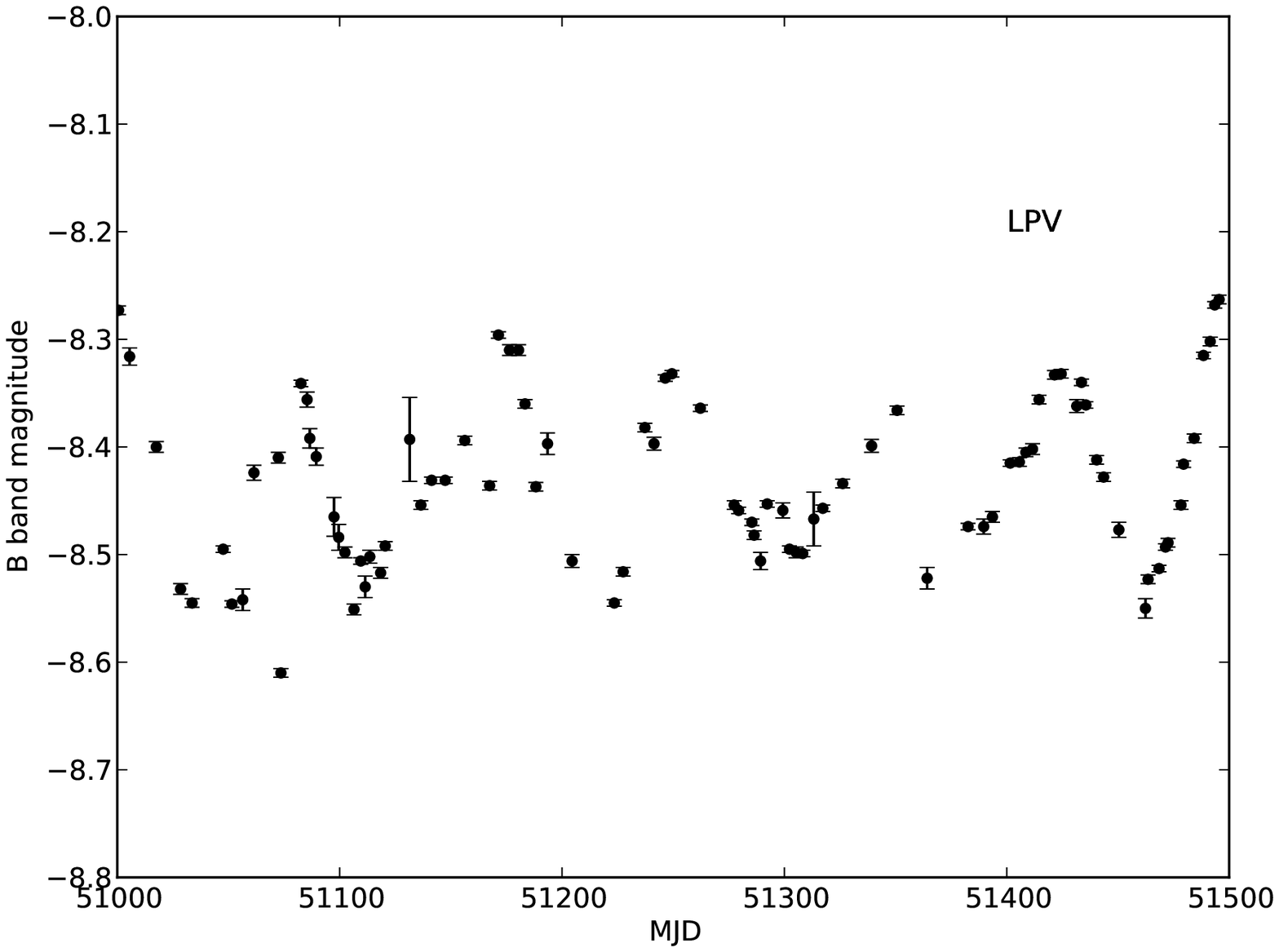}\\

    \includegraphics[width=0.4\textwidth]{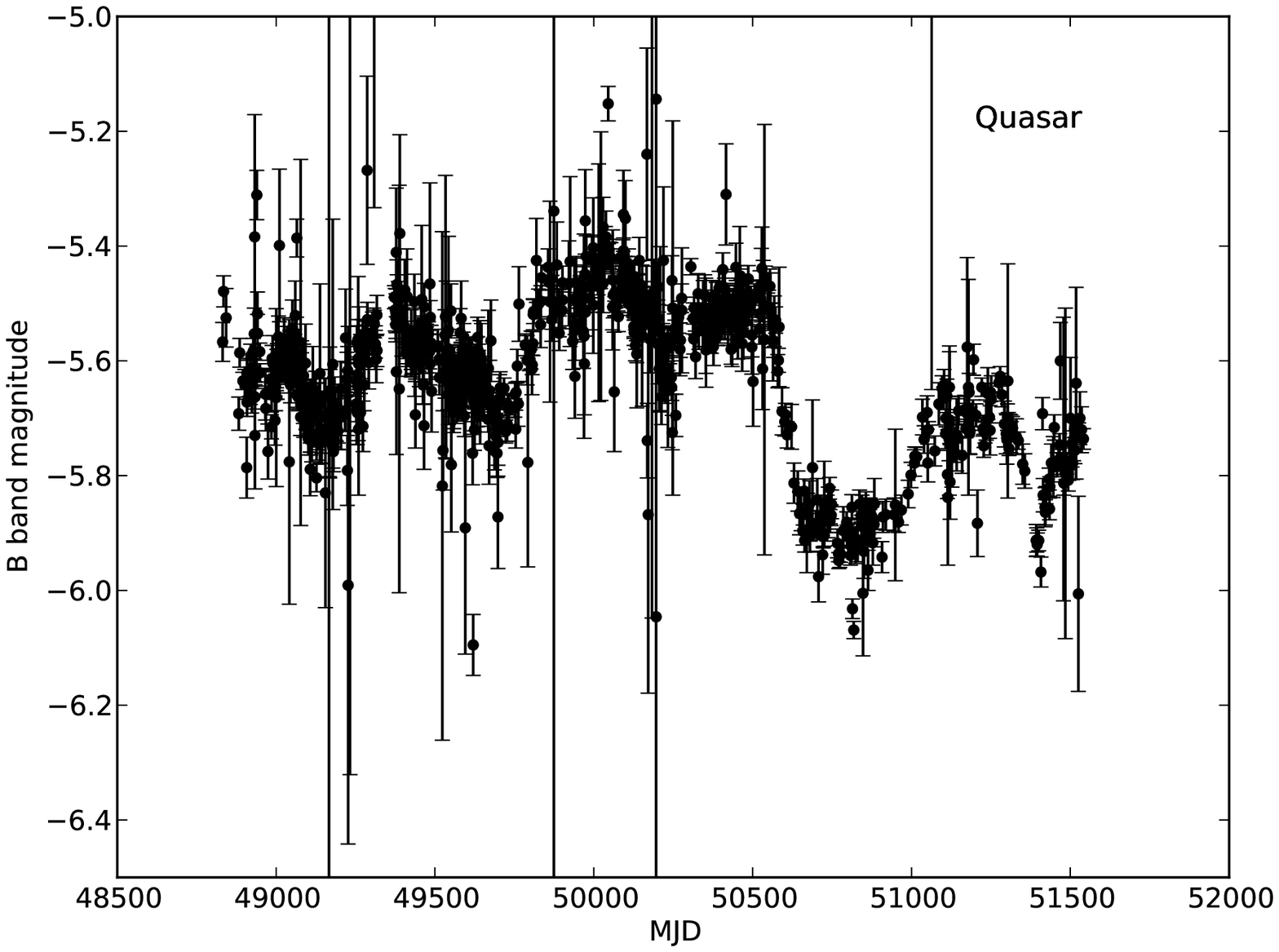}&

    \includegraphics[width=0.4\textwidth]{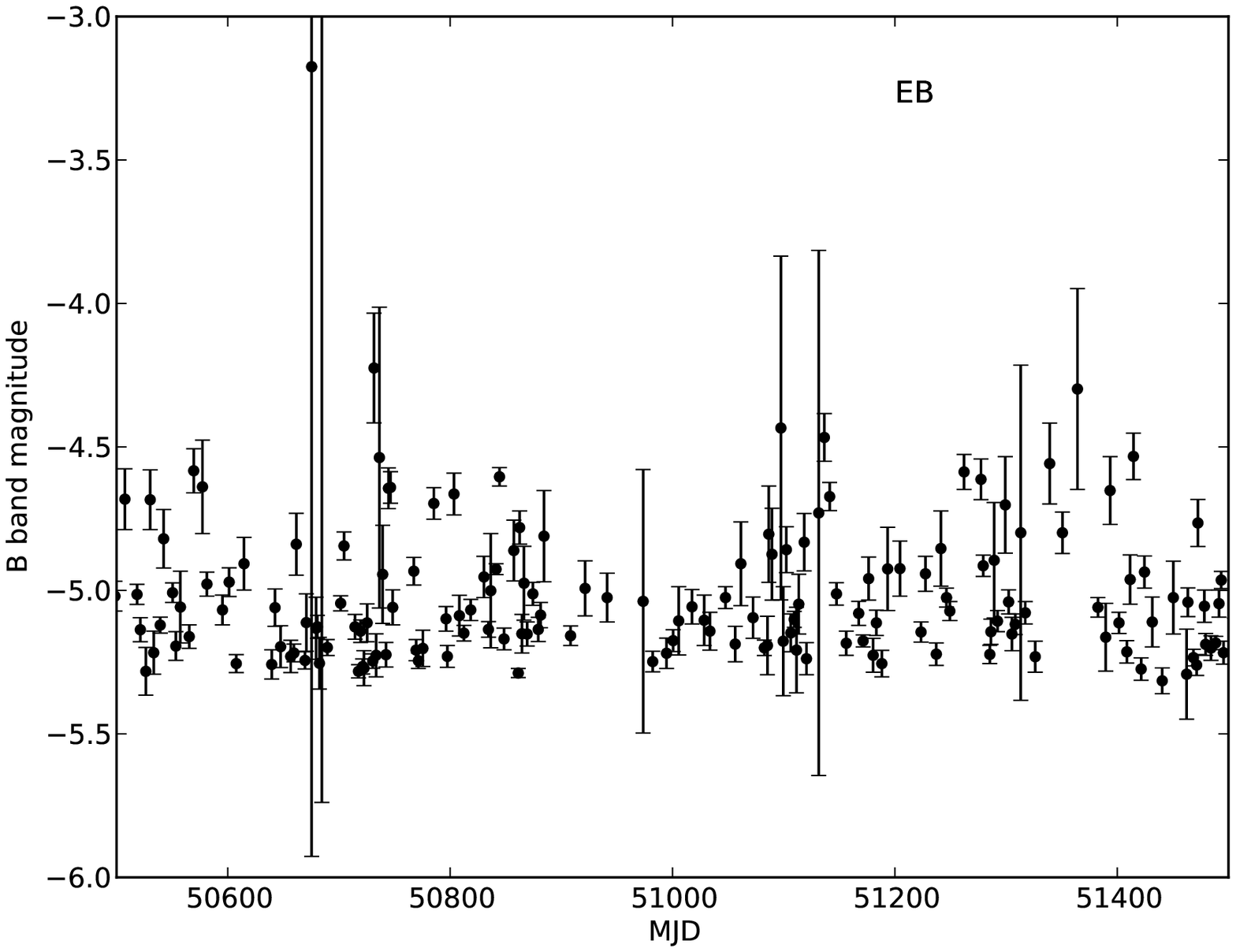}\\
    
    \includegraphics[width=0.4\textwidth]{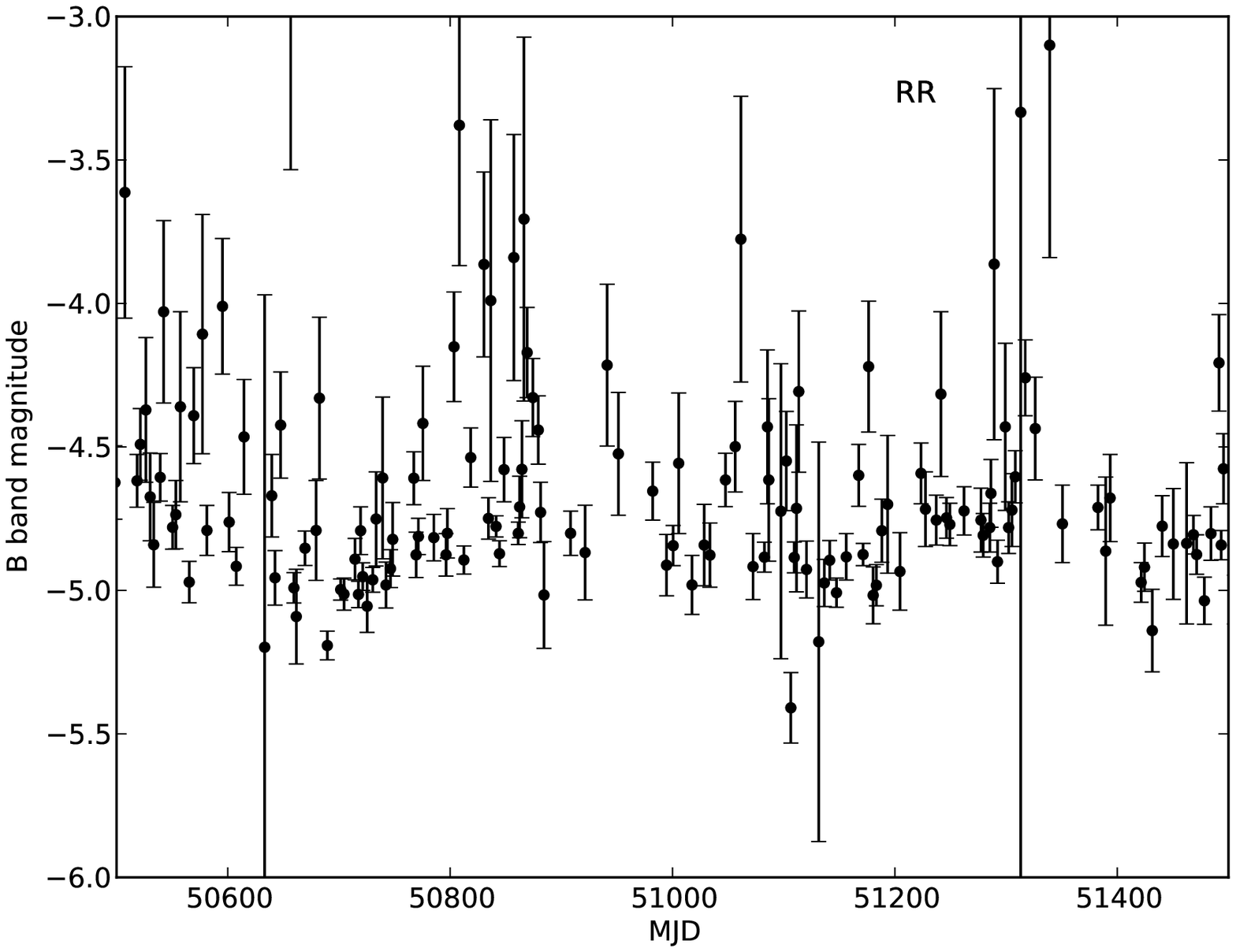}&
    
    \includegraphics[width=0.4\textwidth]{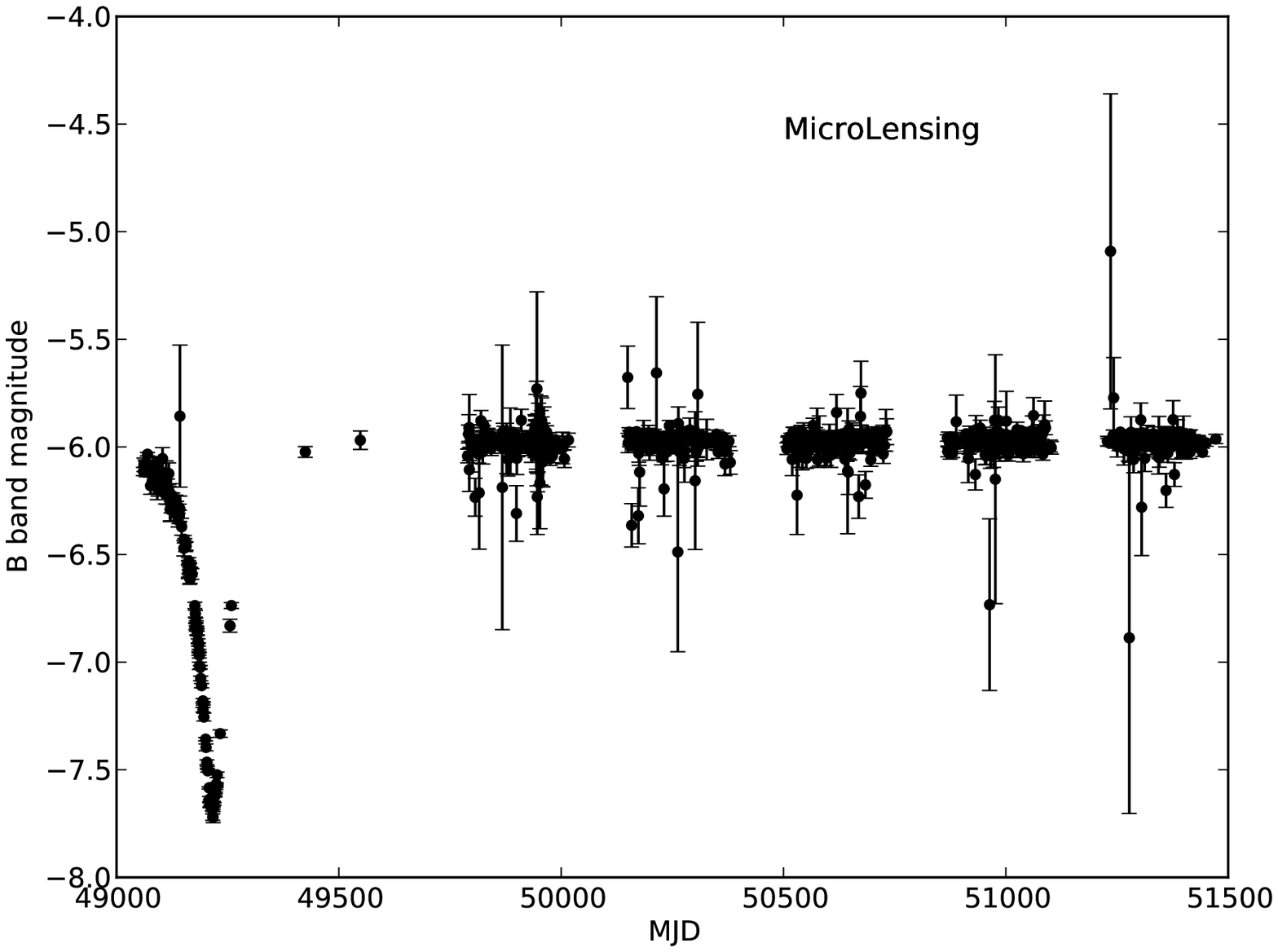}\\

    \includegraphics[width=0.4\textwidth]{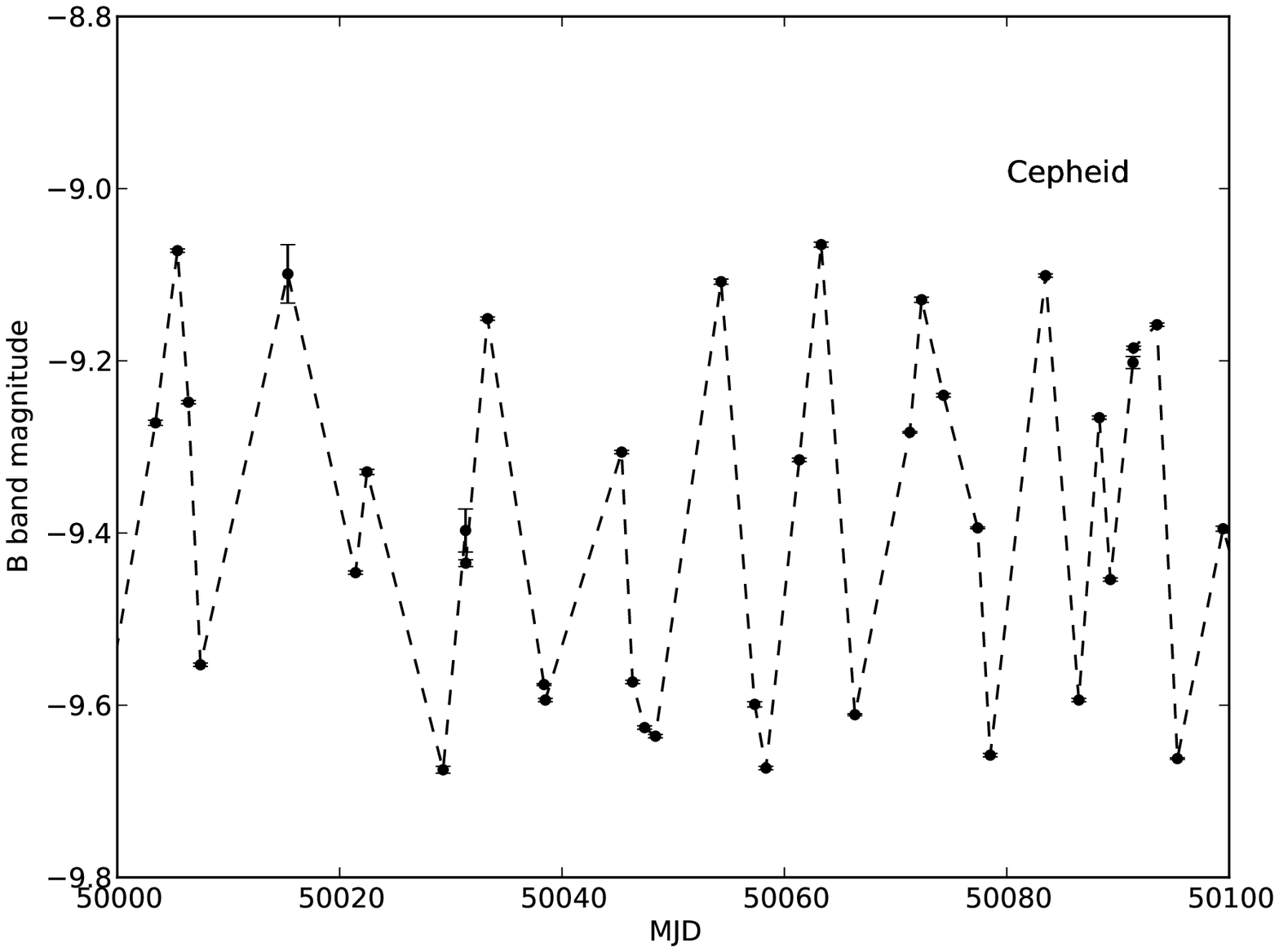}&
        
    \includegraphics[width=0.4\textwidth]{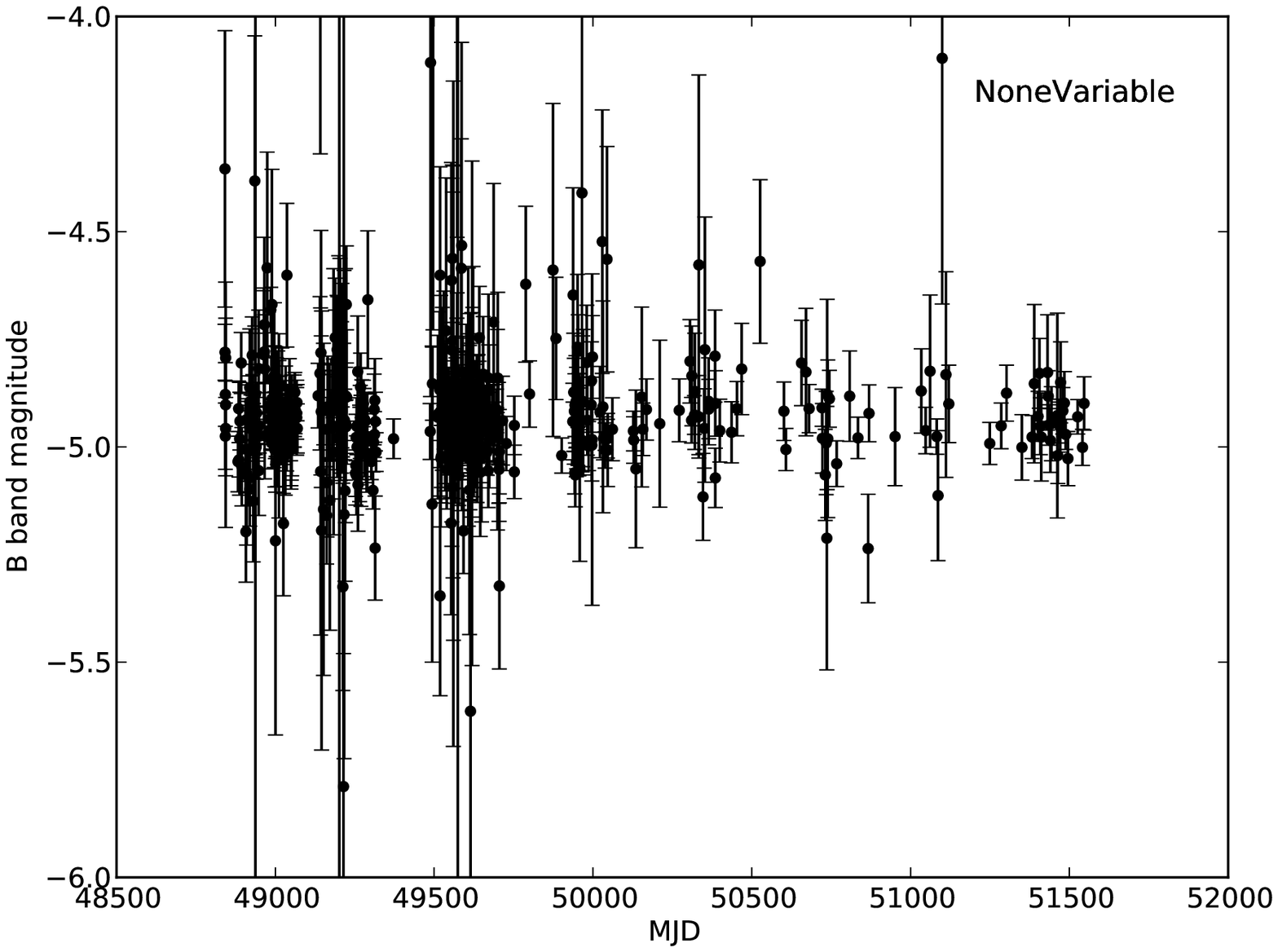}
  \end{tabular}
\caption{Example light curves from MACHO displaying different types of temporal behaviours.}
\label{fig:macho_lcs} 
\end{center} 
\end{figure*}

We demonstrate the effectiveness of both the classifier committee and the hierarchical classifier on optical light curves from the MACHO survey  \cite{alcock2000}.The MACHO observations were carried out between 1992 and 2000 with the 1.27m Great Melbourne telescope at Mount Stromlo Observatory, Australia \cite{alcock2000}. The data come in two photometric bands, the `red' band and the `blue' band. We took a subset of the sample, consisting of 5456 sources from seven source types (number breakdown is provided in Table \ref{tab:macho_sample}). We excluded samples with fewer than 30 observations and source types with less than 50 examples. Example light curves of each source type are shown in Figure \ref{fig:macho_lcs}. Our MACHO dataset is the same as the classified sample used by \cite{kim2011}. The labels of the sources were taken from a number of literature sources \cite{alcock2000, wood2000,keller2002,thomas2005}, and a number of RR Lyraes and Cepheids were identified using a technique described by \cite{wachman2009}. The majority of the light curves span between 2600 and 2800 days of observations. Most light curves have maximum observation gaps of between 50 to 80 days, while on average, the light curves have one observation every two days.

\begin{table}[h]
\caption{MACHO sample by source type} 
\label{tab:macho_sample}
\begin{center} 
\begin{tabular}{lr} 
\hline
Source type &  Number \\
\hline 
\hline 
Non-variable & 3969 \\
Microlensing event & 575 \\
Long Period Variable (LPV) & 361 \\
RR Lyrae star (RR) & 288 \\
Eclipsing binary (EB) & 193 \\ 
Be star & 127 \\
Cepheid variable & 78 \\
Quasar & 58 \\
\hline 
Total & 5456  \\
\hline
\hline 
\end{tabular}
\end{center} 
\end{table}


\section{Light Curve Representation and Classification Algorithm} 
\subsection{Light Curve Representation}\label{s_features} 

Machine learning methods for classification require a structured data set, where each light curve can be represented as \textit{feature vectors} of identical length. Raw light curve observations may not meet this requirement due to differing sampling rates and missing observations.  Hence we must first create homogenised light curve representations for the MACHO data. 

We extracted features from the flux and the flux error measurements of each light curve. These features have been used successfully by \cite{richards2011} and \cite{kim2011} for classifying optical variable sources. The 23 features we used are described below. Here, $x_i$ is the flux measurement and $\sigma_i$ is the error  at each time $i$, $\bar{x}$ is the weighted mean, and $N$ is the number of flux measurements. 
 
\begin{itemize}
  \item {\em Fractional variability}: measures the degree of variability in a set of data points. Fractional variability is similar to the modulation index except that it takes into account the error of each measurement, hence it is more robust against outliers with large error bars. Fractional variability is calculated as: 
  \begin{equation}
  F_{var} = \frac{1}{\bar{x}} \sqrt{ \frac{\sum({x_i-\bar{x})^2} - \sum{\sigma_i^2}} {N}}
  \end{equation}
  
  \item {\em Standard deviation}: calculated as: 
  \begin{equation}
  S = \sqrt{\frac{1}{N-1}\sum(x_i-\bar{x})^2} 
  \end{equation} 
  
  \item {\em Amplitude}: difference between the maximum and minimum flux measurements. 
  
  \item {\em Skew}: A normal distribution has a skew of zero whilst a skew $>0$ means there is more weight in the left tail and vice versa. This is calculated using the Python package \texttt{scipy.stats.skew}\footnote{\url{http://docs.scipy.org/doc/scipy/reference/generated/scipy.stats.skew.html}}. 
  
  \item{\em Beyond1\_std}: the fraction of measurements that are not within $\bar{x}\pm \sigma_i$. 
  
  \item {\em Flux percentile ratios}: We defined a flux percentile
  $F_{n,m}$ to be the difference between the flux values at
  percentiles $n$ and $m$, and use the following flux percentile
  ratios: 
  
   \begin{itemize}
   \item {\em mid-20}: $F_{40,60}/F_{5,95}$
   \item {\em mid-35}: $F_{32.5,67.5}/F_{5,95}$
   \item {\em mid-50}: $F_{25,75}/F_{5,95}$
   \item {\em mid-65}: $F_{17.5,82.5}/F_{5,95}$
   \item {\em mid-80}: $F_{10,90}/F_{5,95}$
   \item {\em percent different flux}: $F_{2,98}$
   \item {\em percent amplitude}: Largest percentage of deviation from median flux.
   \end{itemize} 

   \item {\em Median absolute deviation}: median of the deviations from
       the median value i.e. $median_i(|x_i - median_j(x_j)|)$.
       
   \item {\em Median buffer range percentile}: fraction of observations within 20\% of median flux.
   \item {\em Positive slope trend}: fraction of adjacent flux measurements with positive slope.  

   \item {\em $\eta$}: the ratio of the mean of the square of successive differences to the variance of data points, a useful indicator for the existence of a trend. It is calculated as: 
   \begin{equation}
   \eta = \frac{1}{N-1}\sum\limits_{i=1}^{N-1}\left(\frac{x_{i+1}}{\sigma_{i+1}}\right)^2-\left(\frac{x_{i}}{\sigma_{i}}\right)^2
   \end{equation}
   
   \item {\em Cusum range}: the range of the cumulative sum, which is defined by: 
   \begin{equation} 
   C_k = \frac{1}{N}\sum \limits_{i=1}^{k}\frac{\left(x_i - \bar{x}\right)}{\sigma_i}
   \end{equation} 
   We calculated $C_k$ for all $k$ and the cusum range is $max (C_k) - min(C_k)$. 
   
   \item {\em Lomb-Scargle Periodogram}:
      \begin{itemize}
      \item {\em Period} of the most significant peak 
      \item {\em False alarm probability (FAP)}: high FAP value means the period determined is not statistically significant. 
     \end{itemize} 
     
   \item{\em Autocorrelation features}: Following \cite{kim2011}, we calculated the autocorrelation of the flux measurements and extracted three features as described below. The autocorrelation function typically assumes the even sampling, but it can be modified to accommodate missing measurements \cite{edelson1988}. The autocorrelation function is defined as: 
   
      \begin{equation} 
      AC\left(\tau\right) = \frac{1}{\left(N-\tau\right)\sigma^2} \sum\limits_{i=1}^{N-\tau} \left(x_i-\bar{x}\right)\left(x_{i+\tau}-\bar{x}\right)
      \end{equation} 
   
      \begin{itemize} 
      \item {\em N\_above}: the number of points in $AC(\tau)$ above an empirical boundary as defined by \cite{kim2011}. 
      \item {\em N\_below}: the number of points in $AC(\tau)$ below an empirical boundary as defined by \cite{kim2011}. 
      \item {\em Stetson\_K}: used to characterise the distribution of points in AC($\tau$) \cite{stetson1996}. Stetson K is defined as: 
      \begin{equation} 
      K = \frac{1}{\sqrt{N}} \frac{\sum\limits_{\tau=1}^{N} | AC(\tau)|}{\sqrt{\sum\limits_{\tau=1}^{N}AC(\tau)^2}}
      \end{equation}
      \end{itemize} 
   
   \item {\em B-R}: The weighted mean of the blue magnitudes minus the weighted mean of the red magnitudes. 
   
\end{itemize}

\subsection{Random Forest} 

Our classification algorithm of choice is the random forest (RF), a well-known ensemble classifier \cite{breiman2001} that combine the result of many individual decision tree classifiers \cite{quinlan86}. Decision trees make no assumptions about the parametric distribution of the features and do not require a distance metric. A decision tree is constructed recursively starting with all the training samples at the root node.  The examples are partitioned at each node in the tree via a feature value that produces the highest class label purity in the partitioned sets.  This process continues until all the training samples in a node belong to the same class. RF inserts randomness by only considering a small subset of features (typically only a small fraction of the total number of features) at each node. Each tree in the RF ensemble is trained using a different training sample. The training set is made by bootstrap sampling the original training set (i.e. randomly choosing $S$ samples with replacement). Once the classification ensemble has been trained, it can be used to predict the class of a new sample by getting each decision tree in the ensemble to vote for a class, with the output class being the one with the most votes. 


For the experiments described in this paper, we used 500 trees and then randomly picked the square root of the number of features to use in each node. The RF implementation we used is the \texttt{python} package \texttt{scikit-learn}\footnote{\url{http://scikit-learn.org/}}.  We train the classifier using the features described in Section \ref{s_features}. 

The evaluation metric used throughout this paper is the 10-fold cross-validation accuracy. In 10-fold cross-validation, the labeled data are partitioned into ten stratified\footnote{Stratified cross-validation keeps the proportion of classes the same in all folds. This is especially important for imbalanced datasets to ensure all folds contain similar number of the minority classes.} folds where nine folds are used for training, and the remaining fold is used for testing.  The procedure is repeated ten times where each fold assumes test fold status exactly one time. Accuracy is measured across the test folds where accuracy is defined as the number of samples correctly classified over total number of samples per class. 

\section{Experiments} 
\label{sec:baseline} 

\subsection{Oracle Experiment} 
\label{s_oracle} 

\begin{figure}
    \centering 
        \includegraphics[width=0.5\textwidth]{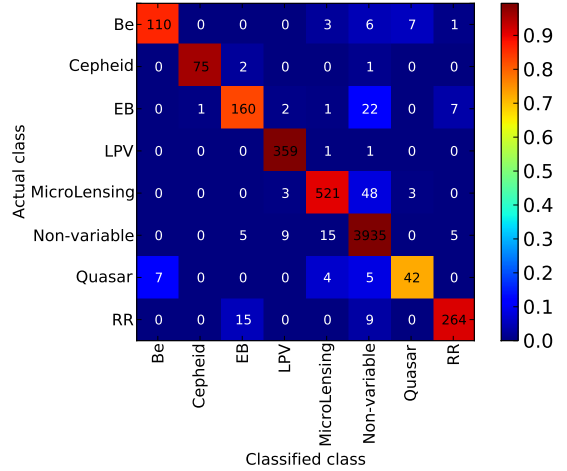}
        \caption{Confusion matrix using the entire time span of light curves in the MACHO data set. The overall accuracy for the MACHO data set is $97\%.$ The colour bar represents the true positive rate.}
        \label{fig:conf_matrix_all}
\end{figure}

The oracle experiment shows the baseline accuracy of our MACHO light curve data using the light curves in their entirety. The 10-fold cross-validation accuracy using the RF classifier on the MACHO dataset is $97\%$. The worst performing class is the quasar class, with an accuracy of $75\%$, which is the same as the performance achieved by \cite{kim2011}. The best performing classes are the LPVs and the non-variables. The MACHO dataset is heavily imbalanced, with non-variables making up $70\%$ of all samples. To maximise overall accuracy, classifiers tend to favour the majority class, therefore it is not surprising that non-variables achieve such good performance. The confusion matrix in Figure \ref{fig:conf_matrix_all} shows that the most frequently confused source classes are microlensing events vs. non-variables, and RR Lyrae stars vs.  eclipsing binaries. Since microlensing events are like non-variables for the parts of the light curves without an event, and RR Lyrae stars and eclipsing binaries have similar periodic behaviour, the results are not surprising. 
    
\subsection{Naive method} 
\label{s:naive}

We now explore methods for classifying light curve data arriving in a stream. The simplest implementation, which we will refer to as the `naive method', is to train a RF classifier using features extracted from the entire length of historical light curves, and then deploy that classifier to new light curves of arbitrary lengths. The advantage of this method is its simplicity -- only one classifier needs to be trained. The uneven sampling of the MACHO light curves means that light curves that span the same number of days do not necessarily have the same number of observations. Gaps of hundreds of days exist in the MACHO light curves and they can degrade the classifier's performance. To minimise the impact of these gaps, we picked the 400-day slice from each light curve with the most number of observations. This implicitly assumes that the light curve is \textit{stationary}, i.e. that its statistical properties do not change when shifted in time and that light curve slices taken at different times provide the same amount of information. 

\begin{figure} 
\begin{center}
\includegraphics[width=0.5\textwidth]{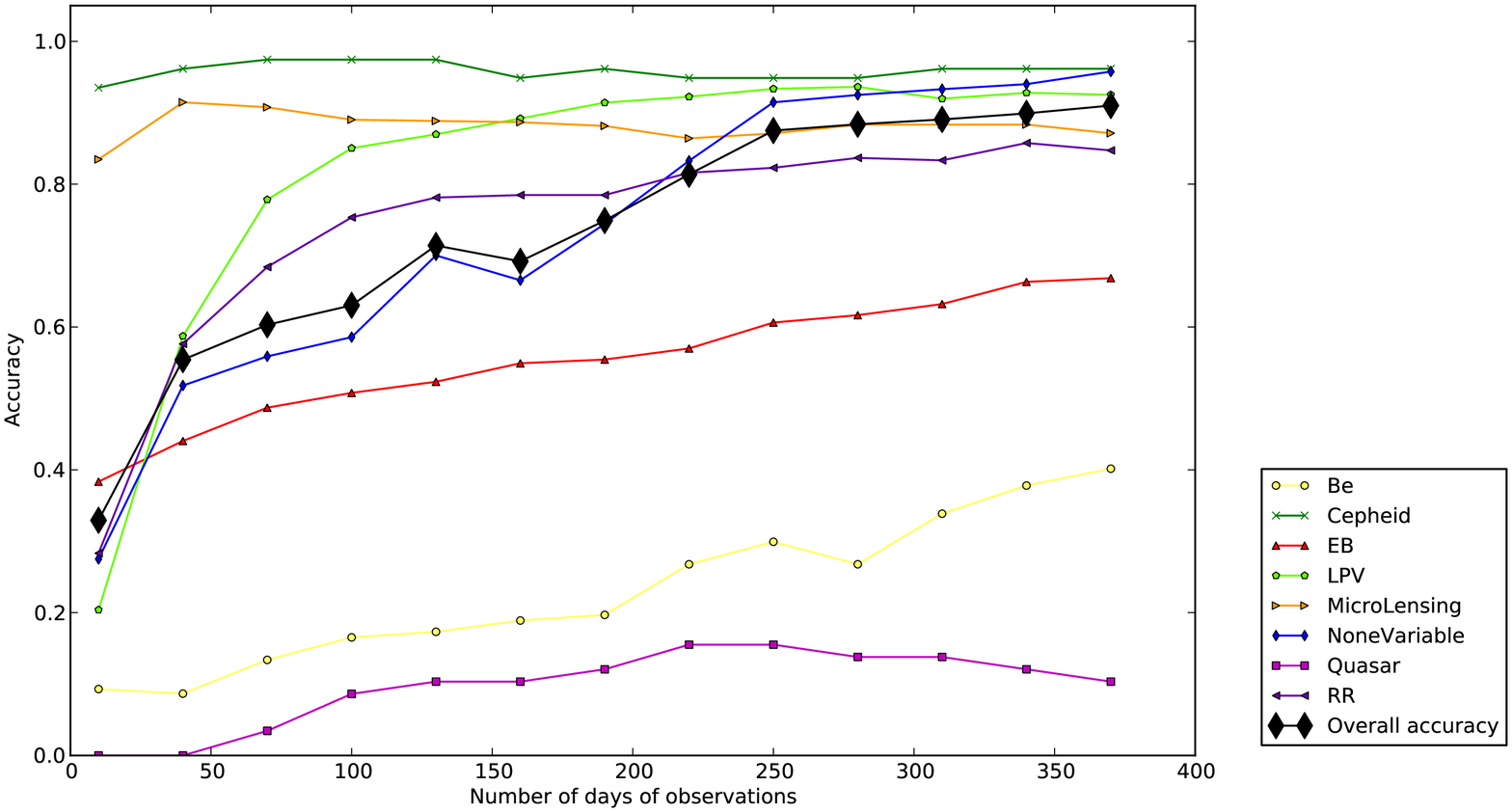} 
\includegraphics[width=0.5\textwidth]{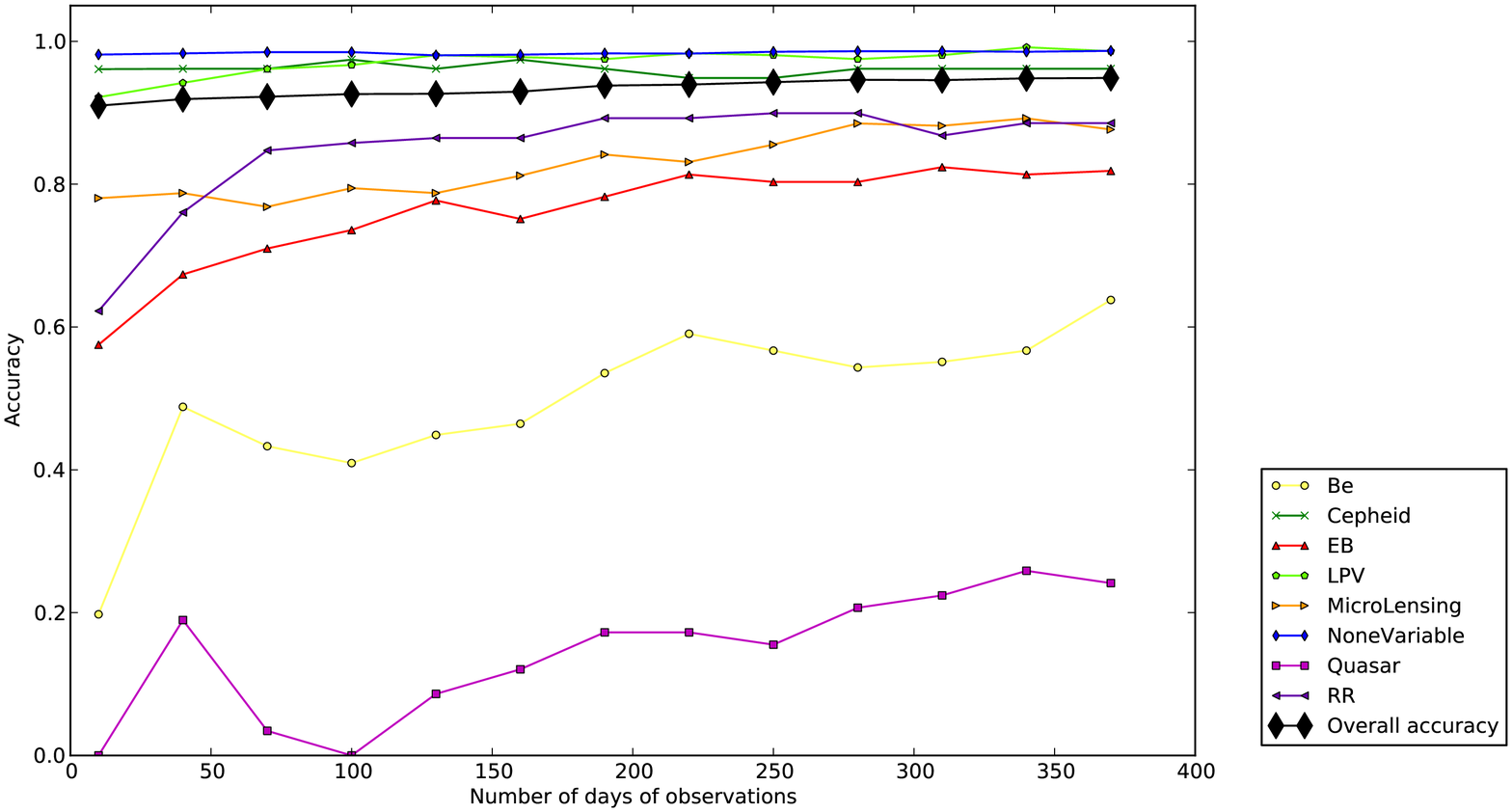} 
\caption{Online classification accuracy of the MACHO dataset with observation window ranging from 10 to 370 days using the naive method (top), the committee method (bottom). } 
\label{fig:macho_results}
\end{center} 
\end{figure}

Figure \ref{fig:macho_results} shows results from 10-fold cross-validation using the naive method. The results show significant divergence in accuracies between the different source types -- with Cepheids and quasars being the best and the worst performer respectively. Accuracies generally increase with the length of the light curves used in the test set. 

\subsection{Committee of Classifiers}
\label{s:committee}

In the naive method, features used in the training set are extracted from a longer light curve than the features used in the test set. This violates the machine learning assumption that the training and the test data are drawn from the same distribution. To remedy the issue of training and test distribution mismatch, we implement a committee of classifiers. We train each classifier in the committee using features extracted from light curves that span different number of days, then in the testing phase, we use the classifier that was trained using the same light curve time span as the test light curves.The final classification decision is made by a single member of the committee, not by an ensemble vote. We perform 10-fold cross-validation on the training set with features extracted using light curves of length between 10 and 370 days. The classification accuracies are significantly higher than those obtained using the naive method and are shown for the MACHO data in Figure \ref{fig:macho_results}. The improvement is most dramatic for non-variables, Cepheids and LPVs, where the accuracies achieved using 70\,days of light curve data are comparable to the oracle classifier's accuracies. Accuracies in other classes, except for quasars and Be stars, also show great improvements in the first 100 days. 


\subsection{Hierarchical Classification}
\label{subs:hierarch}
\begin{figure}
\centering
\includegraphics[width=0.48\textwidth]{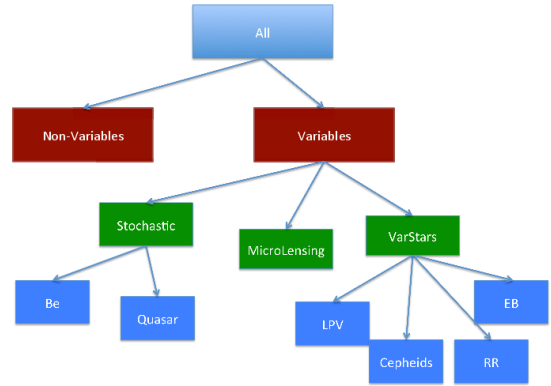}
\caption{Hierarchy of MACHO classes.}
\label{fig:macho_hierarchy}
\end{figure}

Traditional supervised classification trains a classifier to predict a single label from a discrete set of labels. Hierarchical classification predicts a label from an organised hierarchy of classes \cite{silla_survey_2011}, which can be represented as a tree-like structure. A hierarchical classifier that ignores the class hierarchy and only predicts the classes at the leaf nodes is equivalent to a traditional (or `flat') classifier. On the other hand, a hierarchical classifier that trains a different classifier at each level of the hierarchy, in a top-down manner, is analogous to solving a complex problem by breaking it down into a sequence of simpler sub-problems. The advantage of this approach is that a classifier can provide valuable information at each level of the hierarchy, even if the source is ultimately classified incorrectly in the leaf node. In general, misclassifications higher up in the hierarchy are more catastrophic than misclassifications closer to the leaf nodes. 

The first step in hierarchical classification is to create the hierarchical structure, which can either be learned from the data or be created from expert knowledge. For this work, we manually created a hierarchy, shown in Figure \ref{fig:macho_hierarchy}, by grouping the different variability types in the MACHO sources. The top level separation is between variable and non-variable, which is intuitively, the first step in selecting candidates for variability studies. Next, variables are further separated into stochastic variables, variable stars and microlensing events. Stochastic variables have light curves that behave like random walks, such as those of quasars and Be stars. Variable stars are all periodic, and include LPVs, RR Lyrae, EBs and Cepheid variables. Microlensing events have distinctive variability signatures that only occur once. 

Given the hierarchy, the next step is to train the classifiers. We used the method described as {\it local classifier per parent node} by \cite{silla_survey_2011}. In this method, each node is treated as a different sub-problem and it works by training a multi-class classifier for each parent node (i.e. a node with children). For our MACHO hierarchy, there are four parent nodes --  All, Variables, Stochastic and VariableStars, which means four classifiers are trained, and each are trained using only the sources belonging to the class of the node. Once the classifiers are trained, an unknown source can be classified by propagating it down the hierarchy of classifiers until it reaches a leaf node. 

\begin{figure} 
\begin{center}
\includegraphics[width=0.5\textwidth]{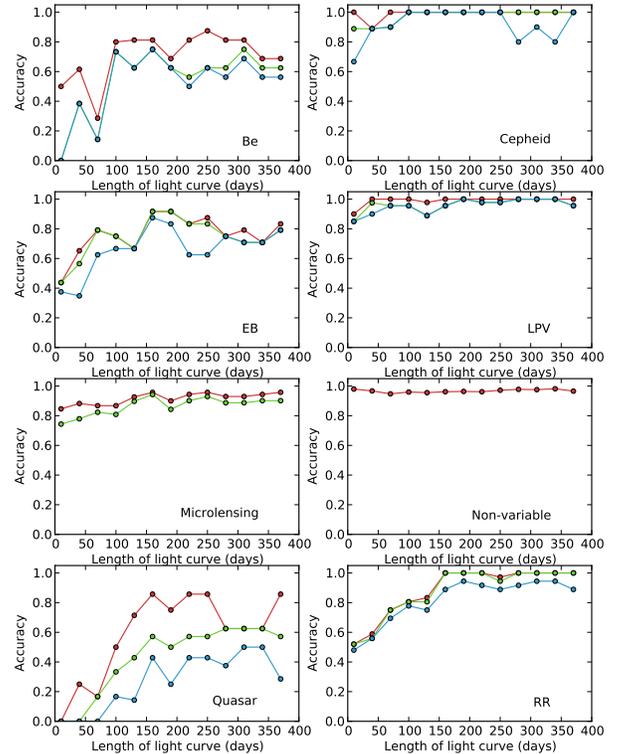} 
\caption[Online classification accuracy from 10 to 370 days up to each of three levels of the hierarchy]{Online classification accuracy from 10 to 370 days up to each of three levels of the hierarchy as illustrated in Figure \ref{fig:macho_hierarchy}.} 
\label{fig:macho_hierarch_acc}
\end{center} 
\end{figure}

The results of our hierarchical classification experiments are shown in Figure \ref{fig:macho_hierarch_acc}. The lines are colour-coded in the same way as the diagram in Figure \ref{fig:macho_hierarchy}. The red line shows the accuracy of classifying sources as variables or non-variables; the green line shows the accuracy for variable stars, microlensing events and stochastic variables; the blue line shows the accuracy for Be, quasars, LPVs, Cepheids, RR Lyrae and EB. Here, accuracy means the fraction of sources of that class that are classified correctly at that level of the hierarchy. To reach the non-variables leaf node only requires one classifier, hence there is only one line in the accuracy plot for the non-variables. Similarly, only two classifiers are needed to get to the leaf node of the microlensing class. For the Be and quasar classes, accuracies with only 130 days of data are better than the performance achieved using the committee method, and accuracies differ significantly at the different levels of the hierarchy.  

One of the disadvantages of hierarchical classification is that an error at the top of the hierarchy always propagates to the lower levels of the hierarchy. For example, if a quasar is misclassified as a non-variable at the top of the hierarchy, it will be wrong on all levels of the hierarchy. Our results show that the hierarchical classifier can attain accuracies at the leaf node levels that are similar to the committee method, and higher accuracies at the higher levels of the hierarchy.  
 
\begin{figure} 
\begin{center}
\includegraphics[width=0.5\textwidth]{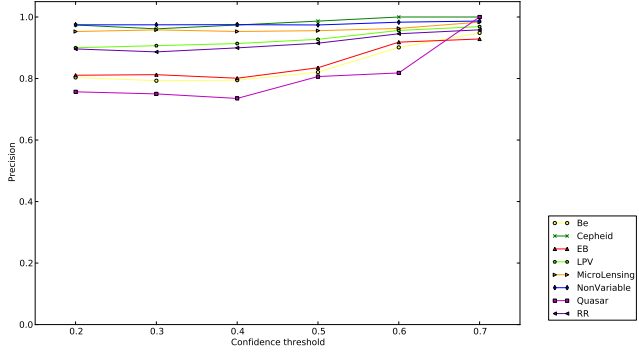} 
\caption{Precision (defined as true positives divided by the sum of true positives and false positives) as a function of the confidence threshold.} 
\label{fig:precision}
\end{center} 
\end{figure}

\begin{figure} 
\begin{center}
\includegraphics[width=0.5\textwidth]{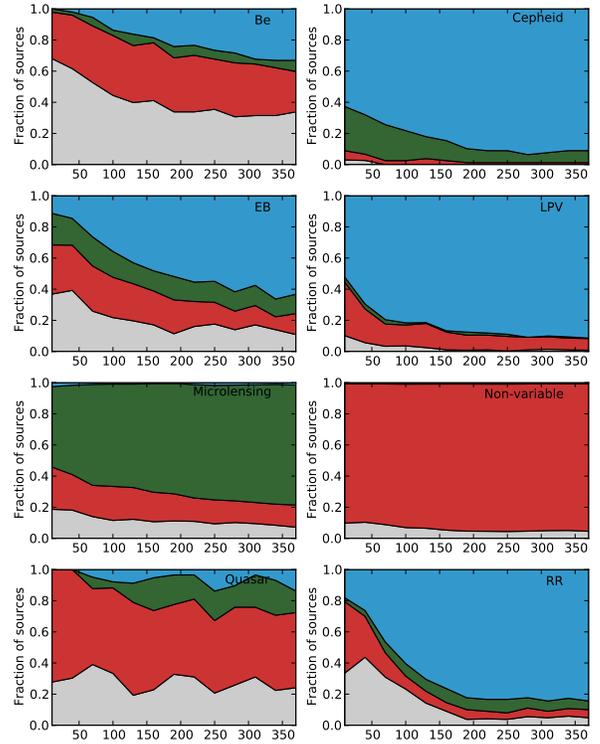} 
\caption{Given the confidence threshold of 0.7, the proportion of sources that terminates at the different levels of the hierarchy illustrated in Figure \ref{fig:macho_hierarchy} from 10 to 370 days. } 
\label{fig:macho_hierarch_crit}
\end{center} 
\end{figure}

Hierarchical classification can be used in the context of online classification by classifying a source at a level of the hierarchy that is commensurate with the amount of information available. This can done by stopping the classification at a level of the hierarchy where the classification confidence falls below a user defined level. One proxy for classification confidence is the classification probability given by the RF classifier. Figure \ref{fig:precision} shows that the precision (defined as true positives divided by the sum of true positives and false positives) increases with increasing confidence threshold. We experimented with setting the confidence threshold to 0.7 such that if a source is not classified with a classification probability of at least 0.7, then the classifier will not proceed to the next level of the hierarchy. Figure \ref{fig:macho_hierarch_crit} shows the proportion of sources that have terminated at each level of the MACHO hierarchy. The grey bands represent sources that cannot be confidently classified as variables or non-variables. Be stars, EBs and quasars, the worst performing classes as shown in Figure \ref{fig:macho_hierarch_acc}, have the highest proportion of sources in the grey bands. The red bands represent sources whose classification stopped at the variables or non-variables level; the green bands represent sources that stopped at the level of micro-lensing events, variable stars and stochastic sources levels; the blue bands represent sources whose classification proceeded all the way to the leaf node. The level at which the classification terminates depends on the time span of the light curve. The longer the light curve, the more information it contains and the further down the hierarchy the classifier can confidently attain. 

\begin{figure} 
\begin{center}
\includegraphics[width=0.5\textwidth]{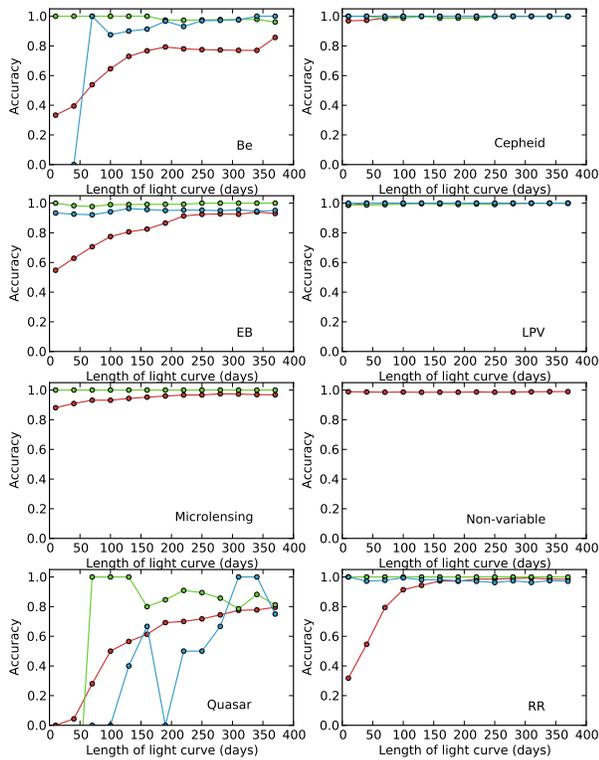} 
\caption{Given the confidence threshold of 0.7, the classification accuracy at each level of the hierarchy. } 
\label{fig:macho_crit_acc}
\end{center} 
\end{figure}

By stopping the classifier before it reaches the leaf node, we can improve the classification accuracy by trading off the level of classification detail. Figure \ref{fig:macho_crit_acc} shows the classification accuracy at each level of the hierarchy, again given the confidence threshold of 0.7. Compared to Figure \ref{fig:macho_hierarch_acc}, only the sources that have proceeded down to a particular level of the hierarchy are used to calculate the accuracy at that level, hence the accuracy at the leaf node level (blue line) is higher than in Figure \ref{fig:macho_hierarch_acc}. 

Hierarchical classification, when used in conjunction with a confidence threshold, allows the user to confidently obtain a classification, albeit at a higher level, when given only a small amount of information. It also allows a different type of classifier to be used at each node. This can mean a different classification algorithm or a different set of features. 

\section{Discussion and conclusion}\label{s_discussion}

In this paper, we explored the feasibility of using light curves for classifying transient and variable sources in an online setting. Although machine-learned classification with light curves has been used in astronomy, classification with a {\em data stream} (online classification) has not received much attention. As we approach the era of large synoptic surveys, online classification will become increasingly relevant. Transient surveys will benefit from having a classification module as part of the processing pipeline.

\begin{table}[h!]
    \centering 
        \begin{tabular}{ lrrrrr } \hline
&  Oracle &  \multicolumn{2}{c}{Naive} &  \multicolumn{2}{c}{Committee }\\
& ${\sim}$2500 days& 160 days & 370 days & 160 days & 370 days \\
\hline 
\hline 
Non-variables & $99\%$ &  $75\%$ & $96\%$ &  $98\%$  & $99\%$  \\
Variables        & $91\%$  & $67\%$ & $80\%$ &   $81\%$  & $86\%$ \\
\hline
Overall           & $97\%$ & $69\%$ & $91\%$  &   $93\%$  & $95\%$ \\ 
\hline 
           
        \end{tabular}
        \caption{Comparison of classification accuracies using different methods}
        \label{tab:comp} 
\end{table}

\begin{table}[h!]
    \centering 
        \begin{tabular}{ lrrrr } \hline
& \multicolumn{2}{c}{Hierarch} & \multicolumn{2}{c}{Hierarch + threshold of 0.7}   \\
 & 160 days & 370 days & 160 days & 370 days \\
\hline 
\hline 
Non-variables     & $97\%$ & $98\%$ & $98\%$ (95\% of srcs) &$99\%$ (96\% of srcs) \\
Variables		   & $82\%$ &  $88\%$ & $99\%$ (68\% of srcs) &$99\%$ (77\% of srcs) \\
\hline
Overall               &  $93\%$ &  $95\%$ & $98\%$ & $99\%$ \\ 
\hline 
           
        \end{tabular}
        \caption{Comparison of leaf node classification accuracies using hierarchical classification.}
        \label{tab:hierarch} 
\end{table}

An online classifier that consists of a committee of classifiers, each learned from light curves of different time spans that match that of the source to be classified, performs significantly better than a classifier that learns from a fully observed light curve. This is shown in the results summarized in Table \ref{tab:comp}. However, the performance differs dramatically across source types. This shows that online classification with light curves has potential for some source classes, but will need to be supplemented with other information to be useful for other source classes. 

We also explored using hierarchical classification with the MACHO dataset. By stopping the classification at the level of the hierarchy where the classification probability is less than the confidence threshold, we can achieve a higher level of accuracy but at a coarser level of classification. The results are summarised in Table \ref{tab:hierarch}. One area to explore is to evaluate the performance of a hierarchical classifier where the hierarchy structure matches that of the follow-up observation requirements. The result of such a classifier would thus be meaningful and can be used for prioritising follow-up observations. More work can also be done on optimising the classifier at each node by selecting the best feature set and algorithm.

\section*{Acknowledgment}

Part of the research described in this paper was carried out at the Jet Propulsion Laboratory under a Research and Technology Development Grant, under contract with the National Aeronautics and Space Administration.  	US Government Support Acknowledged.
Special thanks to Pavlos Protopapas of the Harvard Smithsonian-Center for Astrophysics for providing the labeled MACHO data.



\bibliographystyle{IEEEtran}
\bibliography{journals,onlineclass,optical_surveys,classification}

\begin{thebibliography}{10}
\providecommand{\url}[1]{#1}
\csname url@samestyle\endcsname
\providecommand{\newblock}{\relax}
\providecommand{\bibinfo}[2]{#2}
\providecommand{\BIBentrySTDinterwordspacing}{\spaceskip=0pt\relax}
\providecommand{\BIBentryALTinterwordstretchfactor}{4}
\providecommand{\BIBentryALTinterwordspacing}{\spaceskip=\fontdimen2\font plus
\BIBentryALTinterwordstretchfactor\fontdimen3\font minus
  \fontdimen4\font\relax}
\providecommand{\BIBforeignlanguage}[2]{{%
\expandafter\ifx\csname l@#1\endcsname\relax
\typeout{** WARNING: IEEEtran.bst: No hyphenation pattern has been}%
\typeout{** loaded for the language `#1'. Using the pattern for}%
\typeout{** the default language instead.}%
\else
\language=\csname l@#1\endcsname
\fi
#2}}
\providecommand{\BIBdecl}{\relax}
\BIBdecl

\bibitem{tyson_large_2003}
J.~A. {Tyson}, ``{Large Synoptic Survey Telescope: Overview},'' in
  \emph{Society of Photo-Optical Instrumentation Engineers (SPIE) Conference
  Series}, ser. Society of Photo-Optical Instrumentation Engineers (SPIE)
  Conference Series, J.~A. {Tyson} and S.~{Wolff}, Eds., vol. 4836, Dec. 2002,
  pp. 10--20.

\bibitem{taylor2013}
A.~R. {Taylor}, ``{The Square Kilometre Array},'' in \emph{IAU Symposium}, ser.
  IAU Symposium, vol. 291, Mar. 2013, pp. 337--341.

\bibitem{eyer_automated_2004}
L.~Eyer and C.~Blake, ``Automated classification of variable stars for all-sky
  automated survey 1-2 data,'' \emph{MNRAS}, vol. 358, no.~1, pp. 30--38, 2005.

\bibitem{debosscher_automated_2007}
J.~{Debosscher}, L.~M. {Sarro}, C.~{Aerts}, J.~{Cuypers}, B.~{Vandenbussche},
  R.~{Garrido}, and E.~{Solano}, ``{Automated supervised classification of
  variable stars. I. Methodology},'' \emph{A\&A}, vol. 475, pp. 1159--1183,
  Dec. 2007.

\bibitem{sarro_automated_2008-1}
L.~M. {Sarro}, J.~{Debosscher}, M.~{L{\'o}pez}, and C.~{Aerts}, ``{Automated
  supervised classification of variable stars. II. Application to the OGLE
  database},'' \emph{A\&A}, vol. 494, pp. 739--768, Feb. 2009.

\bibitem{kim2011}
D.-W. {Kim}, P.~{Protopapas}, Y.-I. {Byun}, C.~{Alcock}, R.~{Khardon}, and
  M.~{Trichas}, ``{Quasi-stellar Object Selection Algorithm Using Time
  Variability and Machine Learning: Selection of 1620 Quasi-stellar Object
  Candidates from MACHO Large Magellanic Cloud Database},'' \emph{ApJ}, vol.
  735, p.~68, Jul. 2011.

\bibitem{richards2011}
J.~W. {Richards}, D.~L. {Starr}, N.~R. {Butler}, J.~S. {Bloom}, J.~M. {Brewer},
  A.~{Crellin-Quick}, J.~{Higgins}, R.~{Kennedy}, and M.~{Rischard}, ``{On
  Machine-learned Classification of Variable Stars with Sparse and Noisy
  Time-series Data},'' \emph{ApJ}, vol. 733, p.~10, May 2011.

\bibitem{blomme_automated_2011}
J.~Blomme, L.~M. Sarro, F.~T. {O'Donovan}, J.~Debosscher, T.~Brown, M.~Lopez,
  P.~Dubath, L.~Rimoldini, D.~Charbonneau, E.~Dunham, G.~Mandushev, D.~R.
  Ciardi, J.~De~Ridder, and C.~Aerts, ``Improved methodology for the automated
  classification of periodic variable stars,'' \emph{MNRAS}, vol. 418, pp. 96
  -- 106, Sep. 2011.

\bibitem{long_optimizing_2012}
J.~P. {Long}, N.~E. {Karoui}, J.~A. {Rice}, J.~W. {Richards}, and J.~S.
  {Bloom}, ``{Optimizing Automated Classification of Variable Stars in New
  Synoptic Surveys},'' \emph{PASP}, vol. 124, pp. 280--295, Mar. 2012.

\bibitem{law_palomar_2009}
N.~M. {Law}, S.~R. {Kulkarni}, R.~G. {Dekany}, E.~O. {Ofek}, R.~M. {Quimby},
  P.~E. {Nugent}, J.~{Surace}, C.~C. {Grillmair}, J.~S. {Bloom}, M.~M.
  {Kasliwal}, L.~{Bildsten}, T.~{Brown}, S.~B. {Cenko}, D.~{Ciardi},
  E.~{Croner}, S.~G. {Djorgovski}, J.~{van Eyken}, A.~V. {Filippenko}, D.~B.
  {Fox}, A.~{Gal-Yam}, D.~{Hale}, N.~{Hamam}, G.~{Helou}, J.~{Henning}, D.~A.
  {Howell}, J.~{Jacobsen}, R.~{Laher}, S.~{Mattingly}, D.~{McKenna},
  A.~{Pickles}, D.~{Poznanski}, G.~{Rahmer}, A.~{Rau}, W.~{Rosing}, M.~{Shara},
  R.~{Smith}, D.~{Starr}, M.~{Sullivan}, V.~{Velur}, R.~{Walters}, and
  J.~{Zolkower}, ``{The Palomar Transient Factory: System Overview,
  Performance, and First Results},'' \emph{PASP}, vol. 121, pp. 1395--1408,
  Dec. 2009.

\bibitem{djorgovski_catalina_2011}
\BIBentryALTinterwordspacing
S.~G. Djorgovski, A.~J. Drake, A.~A. Mahabal, M.~J. Graham, C.~Donalek,
  R.~Williams, E.~C. Beshore, S.~M. Larson, J.~Prieto, M.~Catelan,
  E.~Christensen, and R.~H. {McNaught}, ``The catalina real-time transient
  survey {(CRTS)},'' \emph{{arXiv:1102.5004}}, Feb. 2011. [Online]. Available:
  \url{http://arxiv.org/abs/1102.5004}
\BIBentrySTDinterwordspacing

\bibitem{bloom_towards_2008}
J.~Bloom, D.~Starr, N.~Butler, P.~Nugent, M.~Rischard, D.~Eads, and
  D.~Poznanski, ``Towards a real-time transient classification engine,''
  \emph{Astronomische Nachrichten}, vol. 329, no.~3, pp. 284--287, Mar. 2008.

\bibitem{brink_using_2012}
\BIBentryALTinterwordspacing
H.~Brink, J.~W. Richards, D.~Poznanski, J.~S. Bloom, J.~Rice, S.~Negahban, and
  M.~Wainwright, ``Using machine learning for discovery in synoptic survey
  imaging,'' \emph{{arXiv:1209.3775}}, Sep. 2012. [Online]. Available:
  \url{http://arxiv.org/abs/1209.3775}
\BIBentrySTDinterwordspacing

\bibitem{djorgovski_towards_2011}
\BIBentryALTinterwordspacing
S.~G. Djorgovski, C.~Donalek, A.~Mahabal, B.~Moghaddam, M.~Turmon, M.~Graham,
  A.~Drake, N.~Sharma, and Y.~Chen, ``Towards an automated classification of
  transient events in synoptic sky surveys,'' \emph{{arXiv:1110.4655}}, Oct.
  2011. [Online]. Available: \url{http://arxiv.org/abs/1110.4655}
\BIBentrySTDinterwordspacing

\bibitem{alcock2000}
C.~{Alcock}, R.~A. {Allsman}, D.~R. {Alves}, T.~S. {Axelrod}, A.~C. {Becker},
  D.~P. {Bennett}, K.~H. {Cook}, N.~{Dalal}, A.~J. {Drake}, K.~C. {Freeman},
  M.~{Geha}, K.~{Griest}, M.~J. {Lehner}, S.~L. {Marshall}, D.~{Minniti}, C.~A.
  {Nelson}, B.~A. {Peterson}, P.~{Popowski}, M.~R. {Pratt}, P.~J. {Quinn},
  C.~W. {Stubbs}, W.~{Sutherland}, A.~B. {Tomaney}, T.~{Vandehei}, and
  D.~{Welch}, ``{The MACHO Project: Microlensing Results from 5.7 Years of
  Large Magellanic Cloud Observations},'' \emph{ApJ}, vol. 542, pp. 281--307,
  Oct. 2000.

\bibitem{wood2000}
P.~R. {Wood}, ``{Variable red giants in the LMC: Pulsating stars and
  binaries?}'' \emph{PASA}, vol.~17, pp. 18--21, Apr. 2000.

\bibitem{keller2002}
S.~C. {Keller}, M.~S. {Bessell}, K.~H. {Cook}, M.~{Geha}, and D.~{Syphers},
  ``{Blue Variable Stars from the MACHO Database. I. Photometry and
  Spectroscopy of the Large Magellanic Cloud Sample},'' \emph{AJ}, vol. 124,
  pp. 2039--2044, Oct. 2002.

\bibitem{thomas2005}
C.~L. {Thomas}, K.~{Griest}, P.~{Popowski}, K.~H. {Cook}, A.~J. {Drake},
  D.~{Minniti}, D.~G. {Myer}, C.~{Alcock}, R.~A. {Allsman}, D.~R. {Alves},
  T.~S. {Axelrod}, A.~C. {Becker}, D.~P. {Bennett}, K.~C. {Freeman}, M.~{Geha},
  M.~J. {Lehner}, S.~L. {Marshall}, C.~A. {Nelson}, B.~A. {Peterson}, P.~J.
  {Quinn}, C.~W. {Stubbs}, W.~{Sutherland}, T.~{Vandehei}, D.~L. {Welch}, and
  {MACHO Collaboration}, ``{Galactic Bulge Microlensing Events from the MACHO
  Collaboration},'' \emph{ApJ}, vol. 631, pp. 906--934, Oct. 2005.

\bibitem{wachman2009}
G.~Wachman, R.~Khardon, P.~Protopapas, and C.~Alcock, ``Kernels for periodic
  time series arising in astronomy,'' in \emph{Machine Learning and Knowledge
  Discovery in Databases}, ser. Lecture Notes in Computer Science, W.~Buntine,
  M.~Grobelnik, D.~Mladenić, and J.~Shawe-Taylor, Eds.\hskip 1em plus 0.5em
  minus 0.4em\relax Springer Berlin Heidelberg, 2009, vol. 5782, pp. 489--505.

\bibitem{edelson1988}
R.~A. {Edelson} and J.~H. {Krolik}, ``{The discrete correlation function - A
  new method for analyzing unevenly sampled variability data},'' \emph{ApJ},
  vol. 333, pp. 646--659, Oct. 1988.

\bibitem{stetson1996}
P.~B. {Stetson}, ``{On the Automatic Determination of Light-Curve Parameters
  for Cepheid Variables},'' \emph{PASP}, vol. 108, p. 851, Oct. 1996.

\bibitem{breiman2001}
L.~Breiman, ``Random forests,'' \emph{Machine Learning}, vol.~45, pp. 5--32,
  2001.

\bibitem{quinlan86}
J.~R. Quinlan, ``Induction of decision trees,'' \emph{Machine Learning},
  vol.~1, pp. 81--106, 1986.

\bibitem{silla_survey_2011}
C.~N. Silla, Jr. and A.~A. Freitas, ``A survey of hierarchical classification
  across different application domains,'' \emph{Data Min. Knowl. Discov.},
  vol.~22, pp. 31--72, Jan. 2011.

\end{thebibliography}
%

\end{document}